\documentclass[12pt]{article}
\usepackage{epic}
\def\option{\noindent}
\def\vs{\vspace}
\font\cmss=cmss10 at 11pt \font\cmsss=cmss8 at 8pt

\def\inbar{\vrule height1.5ex width.4pt depth0pt}
\def\mininbar{\vrule height.75ex width.3pt depth0pt}
\def\cc{\relax\,\hbox{$\mininbar\kern-.2em{\hbox{\rm\tiny C}}$}}
\def\IZ{\relax\ifmmode\mathchoice
{\hbox{\cmss Z\kern-.4em Z}}{\hbox{\cmss Z\kern-.4em Z}}
{\lower.4pt\hbox{\cmsss Z\kern-.4em Z}}
{\lower1.2pt\hbox{\cmsss Z\kern-.4em Z}}\else{\cmss Z\kern-.4em Z}\fi}
\def\IC{\relax\,\hbox{$\inbar\kern-.3em{\rm C}$}}
\def\IR{\relax{\rm I\kern-.18em R}}
\def\Tr{{\rm Tr}}
\def\mod{{\rm \;mod\;}}
\def\e{{\rm e}}

\newcommand{\Z}{\mathsf{Z}\kern -5pt \mathsf{Z}}
\newcommand{\1}{\mathsf{1}\kern -3pt \mathsf{l}}
\def\ZN{{\Z_N}}
\def\half{ {1\over 2} }
\def\ZCS{  Z_{\rm CS} }
\def\ZqYM{  Z_{\rm qYM} }
\def\twentyfourth{ {1\over 24} }
\def\smhalf{  {\textstyle{1\over 2} }}

\def\sig{\sigma}
\def\Kp{{K'}}
 
\def\W{ W } 
\def\WR{ \W_{R_1 R_2} } 
\def\WRI{ \W_{R_1\ldots R_n} } 
\def\cR{  {\cal R}  }
\def\cM{  {\cal M}  }
\def\cK{  {\cal K}  }
\def\trivial{  {\Sigma_g \times S^1}  }
\def\Seifert{  {\cM_{(g,p)}}}
\def\tc{{\tilde{c}}}
\def\th{{\tilde{h}}}
\def\tN{{\widetilde{N}}}
\def\tR{{\widetilde{R}}}
\def\tS{{\widetilde{S}}}
\def\tT{{\widetilde{T}}}
\def\tcR{ \widetilde{\cR}  }
\def\bk{{\bar k}}
\def\bell{\bar{\ell}}
\def\bR{\bar{R}}
\def\uN{{{\rm u}(N)}}
\def\UN{{{\rm U}(N)}}
\def\UK{{{\rm U}(K)}}
\def\SUN{{{\rm SU}(N)}}
\def\suN{{{\rm su}(N)}}
\def\Uone{{{\rm U}(1)}}
\def\uone{{{\rm u}(1)}}
\def\SUK{{{\rm SU}(K)}}
\def\Spn{{{\rm Sp}(n)}}

\def\Spk{{{\rm Sp}(k)}}
\def\gK{{\widehat{\rm g}_{K}}}
\def\uoneKp{{\widehat{\rm u}(1)_{\Kp}}}
\def\uoneh{{\widehat{\rm u}(1)_{N(K+N)}}}
\def\uoneth{{\widehat{\rm u}(1)_{K(K+N)}}}
\def\uNKh{{\widehat{\rm u}(N)_{K,N(K+N)}  }}
\def\uKNh{{\widehat{\rm u}(K)_{N,K(K+N)}  }}
\def\suNK{{\widehat{\rm su}(N)_K}}
\def\spnk{{\widehat{\rm sp}(n)_k}}
\def\spkn{{\widehat{\rm sp}(k)_n}}
\def\suKN{{\widehat{\rm su}(K)_N}}
\def\one{  {\vcenter  {\vbox  
              {\hrule height.4pt
               \hbox {\vrule width.4pt  height3pt  
                      \kern3pt 
                      \vrule width.4pt height3pt}
               \hrule height.4pt}
                         }
              }
           }
\def\be{\begin{equation}}
\def\ee{\end{equation}}
\def\bea{\begin{eqnarray}}
\def\eea{\end{eqnarray}}

\def\theequation{\thesection.\arabic{equation}}

\begin{document}
\bibliographystyle{bst}

\begin{flushright}
{\tt hep-th/0703089}\\
BRX-TH-582\\
BOW-PH-139\\
\end{flushright}
\vspace{30mm}

\vspace*{.3in}

\begin{center}
{\Large\bf\sf  
Level-rank duality of the $\UN$ WZW model,
\\
Chern-Simons theory, and 2d qYM theory
}
\vskip 5mm Stephen G. Naculich\footnote{Research supported in part
by the NSF under grant PHY-0456944}$^{,a}$
and Howard J.  Schnitzer\footnote{Research supported in part 
by the DOE under grant DE--FG02--92ER40706\\
{\tt \phantom{aaa} schnitzr@brandeis.edu; naculich@bowdoin.edu}\\
}$^{,b}$

\end{center}

\begin{center}
$^{a}${\em Department of Physics\\
Bowdoin College, Brunswick, ME 04011}

\vspace{.2in}

$^{b}${\em Theoretical Physics Group\\
Martin Fisher School of Physics\\
Brandeis University, Waltham, MA 02454}
\end{center}
\vskip 2mm

\begin{abstract}

We study the WZW, Chern-Simons, and 2d qYM theories with
gauge group $\UN$. 
The $\UN$ WZW model is only well-defined for odd level $K$,
and this model is shown to exhibit level-rank duality
in a much simpler form than that for $\SUN$.
The $\UN$ Chern-Simons theory on Seifert manifolds exhibits
a similar duality, distinct from the level-rank duality 
of $\SUN$ Chern-Simons theory on $S^3$.
When $q=\e^{2\pi i/(N+K)}$, 
the observables of the 2d $\UN$ qYM theory can be expressed 
as a sum over a {\it finite} subset of $\UN$ representations. 
When $N$ and $K$ are odd, 
the qYM theory exhibits $N\leftrightarrow K$  duality,
provided $q=\e^{2\pi i/(N+K)}$ and $\theta= 0 \mod  {2 \pi /(N+K) }$.

\end{abstract}

\vfil\break

%\renewcommand{\baselinestretch}{2}
%\small\normalsize

\section{Introduction}
\setcounter{equation}{0}
\label{secintro}

~~~While two-dimensional (Yang-Mills and Wess-Zumino-Witten)
and three-dimensional (Chern-Simons) gauge theories 
have typically been analyzed using simple gauge groups,
for some purposes it is more natural 
to consider these theories with gauge group\break
$\UN~=~[\SUN~\times~\Uone]/\ZN$.

The proposed relation \cite{Ooguri:2004zv} 
between BPS black hole microstates and topological string theory amplitudes
was illustrated in refs.~\cite{Vafa:2004qa,Aganagic:2004js}
by computing bound states of D-branes on a manifold 
that is locally a fibration over a Riemann surface $\Sigma_g$.
The D4-brane worldvolume gauge theory reduces in this case
to a two-dimensional $q$-deformed Yang-Mills (qYM) theory on $\Sigma_g$
with gauge group $\UN$,
which was studied in refs.~\cite{Buffenoir:1994fh}--\cite{Caporaso:2006kk}.
%,%Klimcik:1999kg,
%deHaro:2004wn,%deHaro:2005rz,
%Aganagic:2005dh,
%deHaro:2005rn,
%Arsiwalla:2005jb,%Jafferis:2005jd,Caporaso:2005ta,Caporaso:2005fp,Aganagic:2005wn,deHaro:2006wq,Caporaso:2006gk,Griguolo:2006kp,
%Blau:2006gh,
As we will demonstrate in this paper,
the $\UN$ qYM theory
has the interesting feature---not 
shared by the qYM theory with gauge group $\SUN$---that, 
when 
$q= \e^{2\pi i/(N+K)}$, 
its partition function
is given by a sum restricted to a 
{\it finite} subset of $\UN$ representations,
namely, those with Young tableaux with no more than $K$ columns.
Consequently, the 2d $\UN$ qYM theory exhibits an 
$N \leftrightarrow K$ duality
akin to level-rank duality of WZW 
models \cite{Naculich:1990hg}--\cite{Nakanishi:1990hj}.
%,%Naculich:1990bu,Fuchs:1989rv, 
%Altschuler:1989nm,%Saleur:1990wv, Kuniba:1990im, Walton:1988bs, 
%Mlawer:1990uv,

$\UN$ also plays a more natural role than $\SUN$ 
in large-$N$ dualities of Chern-Simon 
theories \cite{Gopakumar:1998ki},
%Ooguri:1999bv
as pointed out in ref.~\cite{Marino:2001re}.
The $\UN$ Chern-Simons theory also has a natural
realization in terms of free fermions \cite{Douglas:1994ex}.
More recently, Chern-Simons theory on Seifert manifolds 
(circle bundles over $\Sigma_g$) 
has received attention \cite{Beasley:2005vf,deHaro:2005rn,Blau:2006gh}.
In this paper, we will show that 
level-$K$ $\UN$ Chern-Simons theory 
on certain Seifert manifolds
exhibits an $N \leftrightarrow K$ duality
(for odd $N$ and $K$)
that is distinct from the 
$N \leftrightarrow K$ duality that
holds for $\SUN$ Chern-Simons theory 
on $S^3$ \cite{Mlawer:1990uv}, \cite{Camperi:1990dk}--\cite{Labastida:2000yw}.
%,Naculich:1990pa,
%Naculich:1992uf,

Both of the above-mentioned $N \leftrightarrow K$  dualities
flow from the level-rank duality of the Wess-Zumino-Witten model 
with gauge group $\UN$,
which is explored in detail in this paper.
The $\UN$ WZW model has received much less attention than
WZW models based on simple gauge groups,
although its fusion rule algebra was explored in ref.~\cite{Gepner:1992kx}, 
and also in ref.~\cite{Witten:1993xi} in connection
with the quantum cohomology of the Grassmannian.
In several respects, level-rank duality is much simpler
for $\UN$ than for $\SUN$,
and closely resembles that for $\Spn$.
For example, 
level-rank duality of the $\SUN$ WZW model 
involves a 1-1 map between cominimal equivalence classes 
(simple-current orbits) of integrable representations,
whereas for $\UN$ the map is between the representations themselves.

The affine Lie algebra of the $\UN$ WZW model is the quotient 
of $\suNK \times \uoneKp$ by $\ZN$.
To realize the $\UN$ symmetry, 
the levels of the factor groups must be related by $\Kp = N(K+N)$, 
and moreover, 
the level $K$ must be odd \cite{Gepner:1992kx}.
As we will show in section 2,
the primary fields of the $\uNKh$ WZW model (for $K$ odd)
are in one-to-one correspondence with Young tableaux 
with at most $N$ rows and $K$ columns\footnote{Integrable 
representations of $\suNK$ have
Young tableaux with at most $N-1$ rows and $K$ columns.},
and the affine characters of these fields
are infinite sums of characters of $\suNK \times \uoneKp$.
By considering modular transformations 
of these affine characters,
we derive the form of the $S$ and $T$ matrices for $\uNKh$.

In section 3, we demonstrate level-rank duality between the
$\uNKh$ and $\uKNh$ WZW models 
(with $N$ and $K$ odd so that both theories are defined) 
and show the similarity to the duality between $\spnk$ and $\spkn$.
In section 4, 
we use formulas derived from surgery on knots 
to demonstrate level-rank duality of observables of
$\UN$ 
(and also $\Spn$)
Chern-Simons theory on a certain class of Seifert manifolds.
In section 5, 
we show that the partition function of 
2d $q$-deformed $\UN$ Yang-Mills theory
can be expressed,
when 
$q= \e^{2\pi i/(N+K)}$, 
as a sum over a truncated set of $\UN$ representations.
When $K$ is odd, the observables
(partition function, Wilson line expectation values)
can be further expressed in terms of modular
transformation matrices of the $\uNKh$ WZW model.
Consequently, when $N$ and $K$ are odd,
the $\UN$ qYM theory exhibits $N \leftrightarrow K$ duality,
provided $q=\e^{2\pi i/(N+K)}$ and $\theta= 0 \mod  {2 \pi / (N+K) }$.

\section{The $\uNKh$ WZW model} 
\setcounter{equation}{0}

This section is devoted to an analysis of the 
representations of the $\UN$ WZW model,
their affine characters and modular transformation matrices.
The $\UN$ WZW model was previously studied in the context of
its fusion rule algebra in refs.~\cite{Gepner:1992kx,Witten:1993xi}.

The chiral algebra of the $\UN$ WZW model is
the quotient of $\suNK \times \uoneKp$ by $\ZN$.
To realize the $\UN$ symmetry,
the levels of the $\suNK$ and $\uoneKp$ subalgebras
must be related; 
the relation between $K$ and $\Kp$ may be determined
by requiring that the conformal weight of the
representation $(R,Q)$ of $\suNK \times \uoneKp$ 
be proportional to the $\uN$ Casimir (\ref{eq:uncas})
\be
\label{eq:hRQ}
h_{(R,Q)} = h_R + h'_Q 
          =  { \half C_2 (R) \over K+N } +  { Q^2 \over 2K' }
	  =  { \half C_2 (R,Q) \over K+N }
\ee
which implies that \cite{Gepner:1992kx,Witten:1993xi}
\be
\Kp = N(K+N) \,.
\ee
(See the appendix for results and conventions 
for finite-dimensional and affine Lie algebras.)
We will therefore denote the algebra of the $\UN$ WZW model as
\be
\uNKh \equiv [\suNK \times \uoneh]/ \ZN. 
\ee

\vs{.1in}
\noindent{\bf Representations of $\uNKh$ }
\vs{.1in}

\noindent
As in the case of the finite-dimensional $\uN$ algebra
(see the appendix),
representations $(R,Q)$ of $\uNKh$ must satisfy
\be
\label{eq:constraint}
Q = r \mod N
\ee
where $r$ is the number of boxes of the Young tableau associated with $R$,
but in addition we must impose 
\cite{Moore:1989yh,Gepner:1992kx,Witten:1993xi}
the equivalence relation 
\be
\label{eq:idone}
(R,Q) \simeq  (\sig(R), Q+N+K) 
\ee
where $\sig$ is the simple current of $\suNK$, defined in the appendix.
The identification (\ref{eq:idone}) is necessary in order that 
the condition (\ref{eq:constraint}) be preserved
under the modular transformation $\tau \to -1/\tau$.
For the affine characters to be well-defined under $\tau \to \tau + 1$,
moreover, the conformal weights of equivalent representations must differ
by integers.
Using eq.~(\ref{eq:hcomin}), one finds that
\be
h_{(\sig(R), Q+N+K)}- h_{(R, Q)}
=  {2Q-2r+NK+N\over 2N}
=  s  + \smhalf (K+1)
\ee
which is integer-valued if and only if $K$ is odd \cite{Gepner:1992kx}.
Hence, the WZW model with chiral algebra $\uNKh$ is only well-defined 
for odd values of $K$, 
which we henceforth assume\footnote{As another 
example of such a restriction, 
the WZW model with chiral algebra SO(3) 
is only well-defined when the level 
is a multiple of four \cite{Gepner:1986wi,Moore:1989yh,Dijkgraaf:1989pz}.}.

As a result of the equivalence relation (\ref{eq:idone}),
the number of primary fields of $\uNKh$ becomes finite,
so the $\uNKh$ WZW model is a rational conformal field theory.
To determine the number of primaries,
we iterate eq.~(\ref{eq:idone}) $N$ times and use $\sig^N = 1$
to obtain the equivalence
\be
\label{eq:idtwo}
(R,Q) \simeq  (R, Q+N(N+K)) \,.
\ee
Hence $Q$ may be restricted to the range $0 \le Q < N(N+K)$.
{}From this, we determine the number of primary fields of $\uNKh$ to be
\be
\label{eq:number}
\left( N+K-1 \atop K \right) { N(N+K) \over N^2 } =
\left( N+K \atop K \right) 
\ee
where
$\left( N+K-1 \atop K \right)$ 
is the number of integrable highest-weight $\suNK$ representations,\break
$N(N+K)$ is the range of $Q$,
and the constraint (\ref{eq:constraint}) and identification (\ref{eq:idone}) 
each reduce the number of
distinct fields by a factor of $N$. 
(There are no fixed points or short orbits of the equivalence relation.)
The result (\ref{eq:number}) was also obtained 
in the context of Chern-Simons theory in ref.~\cite{Douglas:1994ex}.
We emphasize, however, that this argument is only valid for odd values of $K$,
where it is possible to impose the equivalence relation (\ref{eq:idone}).

As explained in the appendix, 
$\uN$ representations $(R,Q)$ may be characterized by
extended Young tableaux $\cR$ with row lengths 
$\bell_i \in \Z$ $(i = 1, \ldots, N)$.
It is straightforward to prove that, 
within each equivalence class of representations under (\ref{eq:idone}),
there is {\it exactly one} whose extended tableau $\cR$ 
satisfies $0 \le \bell_N \le \cdots \le \bell_1 \le K$.
Hence, 
the primary fields of $\uNKh$
(where $K$ is odd) 
are in one-to-one correspondence with Young tableaux $\cR$ 
with no more than $N$ rows and $K$ columns\footnote{This 
is also the conclusion of ref.~\cite{Caporaso:2006kk},
although the restriction to odd $K$ is not mentioned.}.
The number of such tableaux is $\left( N+K \atop N \right)$, 
in agreement with eq.~(\ref{eq:number}).

\vs{.1in}
\noindent{\bf 
Affine characters and 
modular transformation matrices of $\uNKh$}
 \vs{.1in}

\option 
The affine character for the representation $(R,Q)$ 
of $\suNK \times \uoneKp$ is given by the 
product of eqs.~(\ref{eq:sunchar}) and (\ref{eq:uonechar}),
\be
\label{eq:prodchar}
\chi_{(R,Q)} (\tau) = \chi_R(\tau) \chi'_Q(\tau)\,.
\ee
The primary fields of $\uNKh$, however,
are characterized as equivalence classes of 
representations $(R,Q)$ under the identification (\ref{eq:idone}),
so the affine character\footnote{
Whereas we are considering the basic specialization of the
affine characters, 
Frenkel \cite{Frenkel} considered the {\it principally-specialized} 
characters of $\uN$, but without imposing the identification  (\ref{eq:idone}).
He demonstrated the invariance of these characters 
under transposition of tableaux and $N \leftrightarrow K$.
Historically, this was one of the first manifestations
of level-rank duality.}
of the $\uNKh$ representation $\cR$
is given by the following sum of $\suNK \times \uoneh$ characters 
\bea
\label{eq:unchar}
\chi_{\cR} (\tau)
&=&  \sum_{n\in \Z}
     \chi_{(\sig^n(R),Q + n(N+K))} (\tau)
\nonumber\\
&=&  \sum_{n=0}^{N-1}  
     \sum_{t \in \Z}   
     \chi_{\sig^n(R)} (\tau) \chi'_{Q + (N+K) Nt + (N+K)n} (\tau) 
\nonumber\\
&=&  {1\over \eta(\tau)} \sum_{n=0}^{N-1}  
     \chi_{\sig^n(R)} (\tau)  \Theta_{Q+(N+K)n, N(N+K)/2 } (\tau) 
\eea
where $N(N+K)/2$ is integer-valued since $K$ must be odd,
and 
\be
\Theta_{Q,L}(\tau) 
\equiv 
\sum_{t \in \Z}  \e^{2 \pi i \tau L \left( t + Q/2L \right)^2}  \,.
\ee
Using the modular transformation property of the theta function
\be
{\Theta_{Q,L}\over \eta} \left(-{1\over\tau}\right)
=
\sum_{Q' = 0}^{2L-1}
{\e^{- \pi i Q Q'/L}
\over \sqrt{2L} } 
{\Theta_{Q',L}\over \eta}(\tau)
\ee
together with eqs.~(\ref{eq:sunxf}) and  (\ref{eq:Scomin}),
one finds
\be
\chi_{\cR} (-1/\tau)
=  
	{1\over \eta(\tau)} 
     \sum_{R'} \sum_{Q' = 0}^{N(N+K)-1}
     \left[ \sum_{n=0}^{N-1}   
     \e^{  2 \pi i n (r' - Q') / N }
     \right]
S_{RR'} 
{\e^{- 2 \pi i Q Q'/N(N+K)} \over \sqrt{N(N+K)} } 
     \chi_{R'} (\tau)  \Theta_{Q', N(N+K)/2} (\tau) \,.
\ee
The sum in brackets vanishes unless $Q' = r'$ mod $N$, so we 
may restrict the sum over $Q'$ to obey this constraint
\bea
\label{eq:unxf}
\chi_{\cR} (-1/\tau)
&=&  
{1 \over \eta(\tau)} 
\sum_{R'} 
\sum_{0 \le Q' < N(N+K) \atop Q'=r' \mod N}
N S_{RR'} {\e^{- 2 \pi i Q Q'/N(N+K)} \over \sqrt{N(N+K)} } 
\chi_{R'} (\tau)  \Theta_{Q', N(N+K)/2} (\tau) 
\nonumber\\
&=&  
{1 \over \eta(\tau)} 
\sum_{R'} 
\sum_{0 \le Q' < N(N+K)\atop Q'=r' \mod N}
S_{RR'} {\e^{- 2 \pi i Q Q'/N(N+K)} \over \sqrt{N(N+K)} } 
\sum_{n'=0}^{N-1}
\chi_{\sig^{n'}(R')} (\tau)  \Theta_{Q'+(N+K)n', N(N+K)/2} (\tau) 
\nonumber\\
&=&  
\sum_{\cR'}
S_{RR'} {N \e^{- 2 \pi i Q Q'/N(N+K)} \over \sqrt{N(N+K)} } 
\chi_{\cR'} (\tau)
\eea
where in the second line, we have used 
eq.~(\ref{eq:Scomin})  together with $Q=r$ mod $N$.
The factor of $N$ in the last line arises because each 
primary field $\cR'$
corresponds to $N$ distinct values of $(R', Q')$ satisfying
$0 \le Q' < N(N+K)$ and $Q'=r'$ mod $N$.
{}From eq.~(\ref{eq:unxf}), we read off the modular transformation matrix 
for the $\uNKh$ characters
\be
\label{eq:Sun}
S_{\cR \cR'} =
\sqrt{N\over N+K} S_{RR'} 
\e^{- 2 \pi i Q Q'/N(N+K)}\,.
\ee
(In order not to unduly clutter the notation, we let the subscripts
$R$ or $\cR$ indicate whether the modular transformation 
matrix refers to $\suNK$ or $\uNKh$ respectively.)
In ref.~\cite{Gepner:1992kx}, 
the quantity $S_{\cR \cR'}$  
was obtained in a different way 
by specializing the $\uN$ Weyl character.
This alternative approach allows one to formally define $S_{\cR \cR'} $ 
when $K$ is not odd, 
or even when $K$ is not a real number \cite{Aganagic:2005dh},
but in that case $S_{\cR \cR'} $ does not represent 
the transformation matrix of affine characters under $\tau \to -1/\tau$.
The phase factor in eq.~(\ref{eq:Sun}) was obtained 
in ref.~\cite{Marino:2001re} in the context of $\UN$ Chern-Simons theory,
where $S_{\cR\cR'}$ represents the expectation value of the Wilson 
line of the Hopf link on $S^3$ with canonical framing.

The modular transformation matrix of $\uNKh$ characters (\ref{eq:unchar})
under $\tau \to \tau + 1$ is
\be
\label{eq:Tun}
T_{\cR \cR'} 
= \exp \left[2 \pi i \left(h_{\cR} - {c \over 24}\right) \right] 
\delta_{\cR \cR'}
\ee
where from (\ref{eq:hRQ}) we have
\be
\label{eq:hcR}
h_{\cR} ={ \half C_2 (\cR) \over K+N } 
\ee
and 
\be
\label{eq:unc}
c =  {K(N^2-1) \over K+N} + 1 =  {N(NK+1)\over K+N} \,.
\ee
Note that $T_{\cR \cR'}$ is only well-defined when $K$ is odd because,
when $K$ is even, the conformal weight $h_{\cR}$ 
of the character (\ref{eq:unchar}) 
is not well-defined modulo $\Z$. 

\section{Level-rank duality of the $\uNKh$ WZW model}
\setcounter{equation}{0}

\option
In the  previous section, we showed that the WZW model with affine
Lie algebra $\uNKh$ is only well-defined for odd values of $K$,
for which the equivalence relation (\ref{eq:idone}) may be consistently
imposed.
Throughout this section, 
we restrict ourselves to odd values of $N$ and $K$,
so that both $\uNKh$ and $\uKNh$ are well-defined.

Primary fields of $\uNKh$,
which are characterized (for $K$ odd) by Young tableaux $\cR$ 
with no more than $N$ columns and $K$ rows,
are in one-to-one correspondence\footnote{This is
simpler than level-rank duality between 
the $\suNK$ and $\suKN$ WZW 
models \cite{Naculich:1990hg}--\cite{Nakanishi:1990hj},
%,Altschuler:1989nm,
%Mlawer:1990uv,
in which case the correspondence is one-to-one 
between simple-current orbits of primary fields.}
with the primary fields of $\uKNh$ under transposition of $\cR$.
(The number $\left( N+K \atop K \right)$
of such tableaux is manifestly invariant under $N \leftrightarrow K$.)
In terms of representations of 
$\suNK \times \uoneh$ and $\suKN \times \uoneth$, 
the correspondence is
\be
\label{eq:undual}
\cR  = (R,Q) 
\quad \to \quad \tcR  =
(\sig^{s} (\tR), Q), 
\qquad \qquad s= {Q - r\over N}\,.
\ee
The conformal weights of level-rank-dual primary fields satisfy 
\be
\label{eq:hdual}
h_{\cR}  + \th_{\tcR} = \smhalf Q \,.
\ee
This follows from eqs.~(\ref{eq:hcR}) and (\ref{eq:CcR}) 
and the fact that $T(\cR) = - T(\tcR)$, 
which becomes evident when we rewrite eq.~(\ref{eq:TcR}) as
\be
T(\cR) =  \sum  \bell_i^{\,2}  - \sum  \bk_j^2
\ee
where $\bell_i$ and $\bk_j$ are the row and column lengths,
respectively, of $\cR$. 

{}From eq.~(\ref{eq:unc}), we have that the 
central charges of the level-rank-dual theories  are related by 
\be
c + \tc = NK+1 \,.
\ee
Together with eq.~(\ref{eq:hdual})
this implies that 
\be
\label{eq:Tundual}
\tT_{\tcR\tcR}= (-)^Q \e^{-\pi i (KN+1)/12}  \, T^*_{\cR\cR} 
\ee
where $T$ and $\tT$ are the modular transformation matrices (\ref{eq:Tun}) 
for $\uNKh$ and $\uKNh$ respectively. 

The $S$ modular transformation matrix of $\uKNh$ may be obtained using 
eqs.~(\ref{eq:Sun}) and (\ref{eq:undual}),
\be
\tS_{\tcR\tcR'} 
= \sqrt{K \over K+N}
\tS_{\sig^s(\tR)\sig^{s'}(\tR')} \e^{-2 \pi i Q Q'/K(N+K)} \,.
\ee
Equation (\ref{eq:Scomin}) implies
\be
\tS_{\sig^s(\tR)\sig^{s'}(\tR')} 
= \e^{2 \pi i s Q'/K} \tS_{\tR\sig^{s'}(\tR')}\,, 
\qquad\qquad
\tS_{\tR\sig^{s'}(\tR')}  = \e^{2 \pi i s' r/K} \tS_{\tR\tR'}  \,.
\ee
Finally, under level-rank duality, the $\suNK$ and $\suKN$ 
modular transformation matrices\footnote{
The modular matrix $S$ here differs from that given in 
ref.~\cite{Mlawer:1990uv} by complex conjugation.}
$S$ and $\tS$ are related by \cite{Altschuler:1989nm,Mlawer:1990uv}
\be
\label{eq:Ssundual}
\tS_{\tR\tR'}=  \sqrt{N\over K} \e^{2 \pi irr'/NK} S^*_{RR'}
\ee
which, together with the previous equations, yields
the much simpler level-rank relation
\be
\label{eq:Sundual}
\tS_{\tcR\tcR'} = S^*_{\cR\cR'} 
\ee
between the $\uNKh$ and $\uKNh$ modular transformation matrices $S$ and $\tS$.

Equation (\ref{eq:Sundual})
has the immediate consequence, via the Verlinde formula \cite{Verlinde:1988sn},
that the fusion coefficients of $\uNKh$ are level-rank dual:
\be
\tN_{\tcR\tcR'}^{\ \ \ \tcR^{\prime\prime}} 
=
N_{\cR\cR'}^{\ \ \ \cR^{\prime\prime}} \,.
\ee
Duality of the fusion coefficients also follows, 
as shown in ref.~\cite{Witten:1993xi},
from the symmetry of the Grassmannian 
$G(N,N+K) \simeq G(K,N+K)$.

\vs{.1in}
\noindent{\bf Level-rank duality of the $\spnk$ WZW model}
\vs{.1in}

\option
We note that the level-rank duality of the $\uNKh$  WZW model
is very similar to level-rank duality of 
the $\spnk$ WZW model \cite{Mlawer:1990uv}, 
where $n = $ rank sp($n$).
In the case of $\spnk$, integrable highest-weight representations 
are characterized by Young tableaux $R$ with no more 
than $n$ rows and $k$ columns,
with the level-rank dual representation given by the transpose of $R$.
The relations \cite{Mlawer:1990uv}
between the central charges and conformal weights of $\spnk$ and $\spkn$  
representations
\be
c + \tc = 2 n k, 
\qquad \qquad
h_{R}  + \th_{\tR} = \smhalf r
\ee
and between the modular transformation matrices 
\be
\tS_{\tR\tR'} = S_{RR'} ,\qquad
\tT_{\tR\tR}= (-)^r \e^{-\pi i nk/6} T^*_{RR} 
\ee
are quite similar to those derived above for $\uNKh$.

\section{Chern-Simons theory and level-rank duality}
\setcounter{equation}{0}

Because of the close connection between WZW models in two dimensions 
and Chern-Simons gauge theory in three dimensions \cite{Witten:1988hf}
(see ref.~\cite{Marino:2004uf} for a review),
level-rank duality of WZW models has implications for Chern-Simons
theory as well \cite{Mlawer:1990uv}, \cite{Camperi:1990dk}--\cite{Labastida:2000yw}.
%,Naculich:1990pa,
%Naculich:1992uf,

Since Chern-Simons theory is a topological field theory,
its observables are topological invariants of 
the 3-manifold $\cM$ on which the theory is defined.
The partition function depends only on $\cM$,
the gauge group $G$, the Chern-Simons coupling $K$,
and a choice of framing of the manifold \cite{Witten:1988hf}.
Other gauge-invariant observables of this theory are 
expectation values of path-ordered integrals 
$\Tr_R P  \exp( \oint_\cK A )$
around a closed path (knot) $\cK$ in $\cM$,
with the trace taken in some irreducible representation $R$ of $G$. 
More generally, one may consider Wilson lines on a link 
comprising several, possibly interconnected, knots,
each associated with its own representation.
Expectation values of these Wilson lines (together with a choice of framing)
are topological invariants of the knots and links in $\cM$, 
and are related to well-known knot polynomials.

Observables of the Chern-Simons theory may be expressed in terms of
quantities of the related $\gK$ WZW model. 
For example, consider Chern-Simons theory on the manifold
$\cM = \trivial$,  
where $\Sigma_g$ is a genus $g$ Riemann surface.
The expectation value of $n$ vertical lines (i.e. wrapped around $S^1$)
in $\trivial$, carrying representations $R_i$,
is \cite{Witten:1988hf,Blau:1993tv,Blau:1993hj,Blau:2006gh}
\be
\label{eq:trivial}
\WRI [\trivial, G, K] = 
\sum_{R}  S^{2-n-2g}_{0R}  \prod_{i=1}^n S_{RR_i}
\ee
where $S_{RR'}$ is the modular transformation matrix 
of the $\gK$ WZW model,
the sum is over integrable representations of $\gK$,
and $0$ denotes the identity representation.
Equation (\ref{eq:trivial}) will serve as a building block for
observables of Chern-Simons theory on manifolds other than $\trivial$.

\vs{.1in}
\noindent{\bf Chern-Simons theory on Seifert manifolds}
\vs{.1in}

\option
More generally, one may 
consider \cite{Beasley:2005vf,deHaro:2005rn,Blau:2006gh} 
Chern-Simons theory on a Seifert manifold 
$\Seifert$, a circle bundle over $\Sigma_g$ with
first Chern class $p$.
(The special case $\cM_{(0,p)}$ is the lens space $S^3/\Z_p$,
a circle bundle over $S^2$.)
The Seifert manifold $\Seifert$ may be obtained from $\trivial$ by surgery,
though different choices of surgery yield different framings of $\Seifert$.
Wilson line expectation values on $\Seifert$
may be computed by the methods of ref.~\cite{Witten:1988hf}.
The link consisting of $n-1$ circles 
carrying representations $R_2,\cdots,R_n$
all linked to a single circle
carrying representation $R_1$
has expectation value \cite{Blau:2006gh}
\be
\label{eq:Seifert}
\WRI [\Seifert, G, K]
=  
\sum_R   K^{(p)}_{R_1 R}  \W_{R R_2 \ldots R_n} [\trivial, G, K] \nonumber\\
=  
\sum_R   (K^{(p)} S)_{R_1 R}  S^{2-n-2g}_{0R}  \prod_{i=2}^n S_{RR_i}
\ee
where the line carrying $R_1$ is in the tubular neighborhood where surgery 
occurs, but the other lines are not, 
and where $K^{(p)}$ is a matrix that depends on the framing
of the manifold.    
For $p=0$ (the trivial circle  bundle),
$ K^{(0)} = 1$, and eq.~(\ref{eq:Seifert}) reduces to eq.~(\ref{eq:trivial}).

When $g=0$ and $p=1$, the Seifert manifold is just $S^3$.
For the canonical framing of $S^3$,  
the matrix $K^{(1)}$ is equal to the modular transformation matrix $S$ 
\cite{Witten:1988hf},
so that eq.~(\ref{eq:Seifert}) with $n=2$ yields the Hopf link on $S^3$
\be
\label{eq:threesphere}
\WR [S^3, G, K] =  S_{\bR_1 R_2}   =  S_{R_1 \bR_2} =  S^*_{R_1 R_2}    
\qquad {\rm (canonical~framing)}
\ee
where $\bR$ is the representation conjugate to $R$,
and we have used $S^2 = C$ (where $C$ is the charge conjugation matrix),
$C^2 =1$, and $S^{-1} = S^\dagger = S^*$.

A different choice of framing for $\Seifert$, 
called Seifert framing, was considered
in refs.~\cite{Beasley:2005vf,deHaro:2005rn,Blau:2006gh}.
For this framing, one has $K^{(p)}  = (TST)^p $
\cite{Blau:2006gh}.
Using the relation $ (ST)^3 = S^2 $, one can rewrite
$K^{(p)}  =   (ST^{-1}S^{-1})^p = S T^{-p} S^{-1} $.
In Seifert framing,  the expectation value (\ref{eq:Seifert}) 
on $\Seifert$ becomes
\be
\label{eq:SeifertSeifert}
\WRI [\Seifert, G, K]
=  \sum_R  T^{-p}_{RR} S^{2-n-2g}_{0R}  \prod_{i=1}^n S_{RR_i}
                             \qquad {\rm (Seifert~framing)}.
\ee
Note that, in Seifert framing, 
the Hopf link expectation value on the three-sphere 
\be
\WR [S^3, G, K]
=  (  S T^{-1}S )_{R_1 R_2}
= \left( T  S T C \right)_{R_1 R_2}
=  T_{R_1 R_1}   S_{R_1 \bR_2 } T_{\bR_2 \bR_2} 
              \quad {\rm (Seifert~framing)}
\ee
differs from the result in canonical framing (\ref{eq:threesphere}) 
by the representation-dependent phase factors in $T$.

\vs{.1in}
\noindent{\bf Level-rank duality of $\SUN$ Chern-Simons 
theory on $S^3$}
\vs{.1in}

\option
We now review level-rank duality of observables of 
the $\SUN$ Chern-Simons theory 
on $S^3$ \cite{Mlawer:1990uv}, \cite{Camperi:1990dk}--\cite{Labastida:2000yw}.
%Naculich:1990pa,
%Naculich:1992uf,
These results can be obtained by using the expression (\ref{eq:threesphere}) 
for the Hopf link expectation value on $S^3$ (with canonical framing)
together with the relation (\ref{eq:Ssundual}) between
the modular transformation matrices of $\suNK$ and $\suKN$.
For example, the partition functions
of level-$K$ $\SUN$  Chern-Simons theory 
and level-$N$ $\SUK$ Chern-Simons theory
on $S^3$ (with canonical framing)
are related by 
\cite{Camperi:1990dk}
\be
\ZCS [S^3, \SUN, K]  =   \sqrt{ K \over N } \ZCS [S^3, \SUK, N]   \,.
\ee
Also, the expectation values of unknots on $S^3$ 
(normalized by the partition function)
of level-rank-dual representations are 
equal 
\be
{ \W_R [ S^3, \SUN, K] \over \ZCS [S^3, \SUN, K] }
=   
{ \W_\tR [ S^3, \SUK, N] \over \ZCS [S^3, \SUK, N] }
\ee
and more generally, one can show by using skein relations
that the normalized expectation value of a knot 
taken in representation $R$ of $\SUN$ 
is equal to the expectation value of the mirror-image knot
taken in representation $\tR$ of $\SUK$
\cite{Camperi:1990dk, Naculich:1990pa}.
Finally, normalized expectation values of Hopf links 
of level-rank-dual theories are related by \cite{Mlawer:1990uv}
\be
{ \WR [ S^3, \SUN, K] \over \ZCS [S^3, \SUN, K] }
=   
\e^{-2 \pi i r_1 r_2/NK}  
{ \W^*_{\tR_1 \tR_2}  [ S^3, \SUK, N] \over \ZCS [S^3, \SUK, N] }
\ee
where complex conjugation corresponds to the mirror-image
Hopf link. 
Note that, for $\SUN$, 
duality only holds because cominimally-equivalent 
unknots have identical expectation values on $S^3$.

\vs{.1in}
\noindent{\bf Level-rank duality of $\UN$ and $\Spn$ 
Chern-Simons theory on Seifert manifolds}
\vs{.1in}

\option
Level-rank duality of $\SUN$ Chern-Simons observables
does {\it not} generally hold on manifolds other than $S^3$, 
because the level-rank-dual map for 
$\suNK$ is between cominimal equivalence classes.
As the simplest example, consider the unknot expectation value 
on $  \cM_{(0,0)} = S^2 \times S^1$:
\be
\W_R [ S^2 \times S^1, \SUN, K] = C_{0 R} .
\ee
This vanishes unless $R$ is the identity representation,
but if $R$ corresponds to the cominimal representation  $\sig(0)$,
the level-rank-dual unknot does not vanish because $\tR$ is
equivalent to the identity representation.

Unlike $\SUN$ Chern-Simons theory,
$\UN$ Chern-Simons theory on $\Seifert$ {\it does} exhibit
level-rank duality (for even $p$), as we will now see.
This is because, as shown in section 3,
level-rank duality of $\UN$ WZW models
involves a map between primary fields,
not cominimal equivalence classes thereof.
(As before, however, we must restrict ourselves
to odd values of $N$ and $K$.)
By combining the link expectation value (\ref{eq:SeifertSeifert})
on a Seifert manifold (in Seifert framing)
with the level-rank-dual relationships 
(\ref{eq:Tundual}) and (\ref{eq:Sundual}) 
for modular transformation matrices of $\uNKh$,
we find 
\bea
\W_{\cR_1 \ldots \cR_n} [ \Seifert, \UN, K]
&=&
\sum_{\cR}  T^{-p}_{\cR\cR} S^{2-n-2g}_{0\cR}  \prod_{i=1}^n S_{\cR \cR_i}
\nonumber\\
&=&
\e^{\pi i p(KN+1)/12}  
\sum_{\tcR}  (-1)^{p Q} 
\tT^{*-p}_{\tcR\tcR} \tS^{*2-n-2g}_{0\tcR}  
\prod_{i=1}^n \tS^*_{\tcR \tcR_i} \,.
\eea
When $p$ is odd, the factor of $(-1)^{Q}$ in
the relation (\ref{eq:Tundual}) introduces an unwanted relative
sign between the terms in the sum.
For even $p$, however, the sign is absent, and we obtain the
level-rank-dual relation 
\be
\label{eq:CSduality}
\W_{\cR_1 \ldots \cR_n} [ \Seifert, \UN, K]
= \e^{\pi i p(KN+1)/12}  
\W^*_{\tcR_1 \ldots \tcR_n} [ \Seifert, \UK, N]
\qquad (p~{\rm even})
\ee
between these link expectation values in
$\UN$ and $\UK$ Chern-Simons theories on a Seifert manifold 
(in Seifert framing).
The cases $n=0$, $n=1$, and $n=2$ of eq.~(\ref{eq:CSduality}) 
yield level-rank-dual relationships between partition functions, 
unknot expectation values, and Hopf link expectation values
on a Seifert manifold
respectively.

A similar relation
\be
\W_{R_1\ldots R_m} [ \Seifert, \Spn, k]
= \e^{ \pi i pnk/6} \, W^*_{\tR_1 \ldots \tR_m} [ \Seifert, \Spk, n]
\qquad (p~{\rm even})
\ee
holds between expectation values of $\Spn$ and $\Spk$ Chern-Simons 
theories on $\Seifert$ (for $p$ even),   
because level-rank duality of $\spnk$ WZW models \cite{Mlawer:1990uv} 
is also a map between representations.

%\vfil\break

\section{2d $\UN$ qYM theory and $N \leftrightarrow K$ duality} 
\setcounter{equation}{0}

\option
The two-dimensional $q$-deformed $\UN$ Yang-Mills theory 
that arises \cite{Vafa:2004qa,Aganagic:2004js}
from computations of bound states of D-branes on a 
fibration over a Riemann surface $\Sigma_g$
has a close connection to both $\UN$ Chern-Simons theory
on a Seifert manifold \cite{Aganagic:2004js,
deHaro:2004id,%deHaro:2004uz,
deHaro:2004wn,%deHaro:2005rz,
Beasley:2005vf,
deHaro:2005rn,
Blau:2006gh}
and to the $\uNKh$ WZW model.
In particular, as we will show,
when $q=\e^{2\pi i/(N+K)}$,
the $\UN$ qYM partition function
can be written as a sum over a finite subset of $\UN$ representations.
Consequently, the $\UN$ qYM theory exhibits
a level-rank-type duality under $N \leftrightarrow K$.

The partition function of the
two-dimensional $q$-deformed $\UN$ Yang-Mills theory on
a genus $g$ Riemann surface $\Sigma_g$ 
is given (up to an overall normalization factor) by
\cite{Aganagic:2004js}
\be
\label{eq:qYM}
\ZqYM [\UN, q, p, \theta]
\sim  \sum_\cR \left( \dim_q \cR \right)^{2-2g} q^{-\half p C_2(\cR)}
\e^{i \theta C_1(\cR) }
\ee
where the sum is over all representations $\cR = (R,Q)$ of $\UN$,
whose Young tableaux have row lengths $\bell_i$ obeying
$-\infty < \bell_N \le \bell_{N-1} \le \cdots \le \bell_1 < \infty$.
In eq.~(\ref{eq:qYM}),
$p$ is related to the coupling in the YM action\footnote{Our $p$
is opposite in sign from that in refs.~\cite{Aganagic:2004js,deHaro:2005rn}
and the same as that in ref.~\cite{Blau:2006gh}.},
$C_2(\cR)$ is the quadratic Casimir (\ref{eq:CcR}),
$C_1(\cR) = Q = \sum_{i=1}^{N} \bell_i$,
and $\dim_q \cR$ is the ``quantum dimension'' of the representation
$\cR$
\be
\label{eq:qdim}
\dim_q \cR  
= \prod_{1 \le i < j \le N}  
{ [ \bell_i - \bell_j + j - i ]_q \over [ j - i ]_q }
\qquad {\rm where} \quad
[x]_q = {  q^{x/2} - q^{-x/2} \over  q^{1/2} - q^{-1/2}  }.
\ee
Letting $m_i = \bell_i -i \in \Z$,
we may rewrite eq.~(\ref{eq:qYM}) as
\bea
\label{eq:qYMi}
&&
\ZqYM [\UN, q, p, \theta]
\sim  \sum_{m_N < \ldots < m_1} 
\left( 
\prod_{1 \le i < j \le N}  
{ [ m_i - m_j ]_q \over [ j - i ]_q }
\right)^{2-2g} 
\times 
~~~~~~~~~~~~~~~~~~
~~~~~~~~~~~~~~~~~~
\\
&& 
~~~~~~~~~~~~~~~~~~~~~~~~~~~~~~~~~~~~~~~~~~
q^{- \half p \sum_i \left[m_i + \half (N+1) \right]^2 
 + \twentyfourth p N(N^2-1) }
\e^{i\theta \sum_i \left[m_i + \half (N+1) \right] } \,.
\nonumber
\eea
The invariance of eq.~(\ref{eq:qYMi}) under permutations of the $m_i$
allows us to extend the sum to all $m_i$ (modulo a factor of $N!$),
omitting the locus where two or more of the $m_i$ are equal.
The resulting expression is actually the result of a path integral
derivation of the 2d qYM partition function \cite{Aganagic:2004js},
in which terms with $m_i =m_j$ 
correspond to singular points that 
give no contribution to the path integral \cite{Blau:1993hj}.

Next, we show that when $q=\e^{2\pi i/(N+K)}$,
with $K$ and $p$ both integer-valued,
the infinite sum over representations in (\ref{eq:qYM})
either vanishes or is proportional to 
a sum over a {\it finite} subset of $\UN$ representations,
namely those whose tableaux have no more than $K$ 
columns.\footnote{When $K$ is not an integer, there is no truncation
of representations \cite{Aganagic:2004js}.}
First, when $q=\e^{2\pi i/(N+K)}$, one has 
$[x]_q = \sin( {\pi x \over N+K} )/\sin( {\pi \over N+K} )$,
so that terms in which two $m_i$'s differ by a multiple of
$N+K$ are singular, 
and should be omitted from the sum (\ref{eq:qYMi})
\cite{Blau:1993hj}.
The resulting partition function is 
\bea
&&\ZqYM [\UN, \e^{2\pi i/(N+K)}, p, \theta]
\sim  {\sum_{m_i \in \Z}}'
\left( \prod_{1 \le i < j \le N}  
{ [ m_i - m_j ]_q \over [ j - i ]_q } \right)^{2-2g} 
\times
~~~~~~~~~~~~
\\
&& 
~~~~~~~~~~~ 
\exp\left\{- {\pi i p\over N+K} \sum_i \left[m_i + \smhalf (N+1) \right]^2 
 + {\pi i p N(N^2-1) \over 12(N+K)} 
+ i\theta \sum_i \left[m_i + \smhalf (N+1) \right] \right\}
\nonumber
\eea
where the prime on the sum denotes the omission of the locus
$m_i = m_j$ mod $N+K$ for all $i \neq j$.

Next, observe that under $m_i \to m_i + N + K$ (with the other $m_j$ fixed),
the summand is multiplied by 
$ \e^{-i \pi p (K+1) +  i(N+K) \theta}$,
provided $p \in \Z$.
If $\theta \neq {\pi p(K+1)\over N+K}$ mod ${2 \pi\over N+K}$,
the resulting infinite sum over phases causes the partition function to vanish.
On the other hand, 
if $\theta = {\pi p(K+1)\over N+K}$ mod ${2 \pi\over N+K}$,
the summand  is invariant under $m_i \to m_i + N + K$,
and so the sum may be restricted 
(modulo an infinite factor)
to the hypercube
$- N \le m_i < K-1 $,
still omitting terms in which any two $m_i$'s are equal. 
Finally, due to the invariance under permutations of $m_i$,
the sum may be restricted to
$ - N \le  m_N < \cdots < m_1 \le K-1 $,
which corresponds to 
$0 \le \bell_N \le \bell_{N-1} \le \cdots \le \bell_1 \le K$.
Hence
\be
\label{eq:qYMI}
\ZqYM \left[\UN, \e^{2\pi i/(N+K)},p, \theta  \right]
= \sum_{0 \le \bell_N \le \cdots \le \bell_1 \le K } 
\left( \dim_q \cR \right)^{2-2g} q^{-\half p C_2(\cR)}
\e^{i \theta C_1(\cR)}
\ee
provided $p \in \Z$ and
$\theta = {\pi p(K+1)\over N+K}$ mod ${2 \pi\over N+K}$.
To summarize, 
when these conditions hold,
the 2d $\UN$ qYM partition function (\ref{eq:qYM}) is proportional to
eq.~(\ref{eq:qYMI}),
which has the same form but is restricted to 
Young tableaux with no more than $K$ columns.
This result holds for both even and odd $K$; 
the allowed set of Young tableaux corresponds 
in the latter case to primary fields of $\uNKh$.

We note that the analogous truncation does {\it not} 
occur in the case of the $\SUN$ qYM partition function even for
$q = \e^{2\pi i/(N+K)}$;
the extra term $-r^2/N$ in the $\SUN$ Casimir (\ref{eq:suncas})
ruins the invariance of the summand under $m_i \to m_i + N + K$
in that case.

As we saw in sec.~2,
when $K$ is odd, 
the set of Young tableaux with up to $N$ rows and $K$ columns
precisely corresponds 
to the set of primary fields of $\uNKh$.
Moreover, the quantities in eq.~(\ref{eq:qYMI}) may be 
expressed in terms of modular transformation matrices 
(\ref{eq:Sun}) and (\ref{eq:Tun}) of $\uNKh$,
namely
$ q^{\half C_2(\cR)} = T_{\cR \cR}/T_{00} $
and 
\bea
\dim_q \cR  
&=&\prod_{1 \le i < j \le N}  
{ [ \bell_i - \bell_j + j - i ]_q \over [ j - i ]_q }
= \prod_{1 \le i < j \le N}  
{ [ \ell_i - \ell_j + j - i ]_q \over [ j - i ]_q }
= \dim_q R  
= \left( S_{0R}\over  S_{00}\right)_{\suNK} 
\nonumber\\
&=& \left( S_{0\cR}\over  S_{00}\right)_{\uNKh}
\eea
where $\ell_i = \bell_i - \bell_N$ are the row lengths of the $\SUN$ 
representation $R$,
and we have used eq.~(\ref{eq:Sun}).
Hence,  when $K$ is odd and $\theta = 2\pi t/(N+K)$ with  $t\in\Z$,
the $\UN$ qYM partition function (\ref{eq:qYMI}) 
may be expressed as
\be
\label{eq:qYMST}
\ZqYM \left[\UN, \e^{2\pi i/(N+K)},p, {2 \pi t \over  N+K} \right]
= {T_{00}^p S_{00}^{2g-2} }
\sum_{\cR ~{\rm integrable}}  
S_{0\cR}^{2-2g} T_{\cR\cR}^{- p } \e^{2 \pi i t Q / (N+K)} 
\qquad (K {\rm ~odd}) \,.
\ee
Finally, as observed in refs.~\cite{Aganagic:2004js,
deHaro:2004wn,%deHaro:2005rz,
Beasley:2005vf,
deHaro:2005rn,
Blau:2006gh},
the $\UN$ qYM partition function 
(for $q = \e^{2\pi i/(N+K)}$ and $\theta=0$)
may be expressed in terms of 
the level-$K$ $\UN$ Chern-Simons partition function 
on $\Seifert$ with Seifert framing  (\ref{eq:SeifertSeifert})
\be
\label{eq:YMCS}
\ZqYM [\UN, \e^{2\pi i/(N+K)},p,0]
= T_{00}^{p} S_{00}^{2g-2}
\ZCS [\Seifert, \UN, K]
\qquad (K {\rm ~odd}) \,.
\ee

\vfil\break
\vs{.1in}
\noindent{\bf $N \leftrightarrow K$ duality
of the 2d $\UN$ qYM theory} 
\vs{.1in}

\option
With the 2d $\UN$ qYM partition function in the form (\ref{eq:qYMST}),
duality under $N \leftrightarrow K$ becomes manifest by using 
eqs.~(\ref{eq:Tundual}) and (\ref{eq:Sundual}):
\be
\label{eq:NKdual}
\ZqYM \left[\UN, \e^{2\pi i/(N+K)},p, {2 \pi t \over N+K }\right]
= \ZqYM^* \left[\UK, \e^{2\pi i/(N+K)}, p, 
{2 \pi t \over  N+K} + \pi p \right]
\quad (N, K~{\rm odd}) \,.
\ee
That is, the 2d $\UN$ qYM partition function 
is the complex conjugate of the 2d $\UK$ qYM partition function, 
provided $q=\e^{2\pi i/(N+K)}$ and $\theta= 0 \mod  {2 \pi /(N+K) }$,
with $N$ and $K$ both odd.
When $p$ is odd, $\theta$ must be shifted in order 
to compensate for the representation-dependent minus sign 
in (\ref{eq:Tundual}),
but when $p$ is even, no shift is necessary.

It has been shown \cite{deHaro:2005rn} 
that a certain class of Wilson line observables in qYM theory 
(for $q= \e^{2\pi i/(N+K)}$) 
are proportional to the expectation values of links 
of Chern-Simons theory on $\Seifert$. 
For $K$ odd, 
these Wilson line observables in $\UN$ qYM theory 
may be expressed in terms of modular transformation matrices of $\uNKh$:
\be
\W_{\cR_1\ldots \cR_n} 
\left[\UN, \e^{2\pi i/(N+K)},p, {2 \pi t \over  N+K} \right]
\sim 
\sum_{\cR ~{\rm integrable}}  
S_{0\cR}^{2-2g-n} T_{\cR\cR}^{- p } \e^{2 \pi i t Q / (N+K)} 
 \prod_{i=1}^n S_{\cR\cR_i}
\quad (K {\rm ~odd})\,.
\ee
(The sum over all $\uN$ representations 
restricts to a sum over a finite subset of representations 
under the same conditions as before.)
Once again,
eqs.~(\ref{eq:Tundual}) and (\ref{eq:Sundual})
can be used to show that these observables, 
like the partition function, 
exhibit $N \leftrightarrow K$ duality (\ref{eq:NKdual}). 

\section{Conclusions}
\setcounter{equation}{0}
\label{secconcl}

In this paper we have examined three different 
exactly-soluble gauge theories in two and three dimensions 
with gauge group $\UN$, and
the $N \leftrightarrow K$ dualities that these theories exhibit.

First, we examined the $\UN$ WZW model, whose 
affine Lie algebra is 
$\uNKh \equiv [\suNK \times \uoneh]/ \ZN$.
The primary fields of this theory are given by
equivalence classes of representations
of $\suNK \times \uoneh $.
These equivalence classes are only defined when $K$ is odd,
and are in one-to-one correspondence with Young tableaux 
with up to $N$ rows and $K$ columns.
By explicitly considering the affine characters of the
primary fields, we derived the modular transformation matrices
of the $\uNKh$ WZW model.
We then showed that the $\uNKh$ and $\uKNh$ WZW models are 
level-rank dual (when $N$ and $K$ are odd), 
with a one-to-one correspondence between
the primary fields of these theories.
This is simpler than the level-rank duality between 
$\suNK$ and $\suKN$ WZW models, 
where the correspondence is between 
the simple-current orbits of primary fields.

Next we considered $\UN$ Chern-Simons theories on Seifert manifolds:
circle bundles over $\Sigma_g$ with first Chern class $p$.
The partition function and knot and link expectation values
of this theory were expressed (using surgery) in terms of the
$\uNKh$ modular transformation matrices. 
The level-rank duality of the latter
was used to show the $N \leftrightarrow K$ duality of the
Chern-Simons observables, for $N$ and $K$ odd, 
and $p$ even.  
A similar result was shown for $\Spn$ Chern-Simons theory
on Seifert manifolds.

Finally, we considered two-dimensional $q$-deformed $\UN$ 
Yang-Mills theory.   
We showed that when $q=\e^{2\pi i/(N+K)}$,
the partition function may be expressed 
in terms of a finite sum over $\uN$ representations.
This result, together with the level-rank duality of the
$\uNKh$ WZW  model,  was then used to show that 
the 2d $\UN$ qYM theory exhibits a level-rank-type duality 
under $N \leftrightarrow K$ (for $N$ and $K$ both odd),
provided $q=\e^{2\pi i/(N+K)}$ and $\theta= 0 \mod  {2 \pi /(N+K)}$.

The $N \leftrightarrow K$ duality of $q$-deformed YM theory 
discussed in this paper is not immediately relevant to 
the counting of bound states of BPS black holes in string theory 
because in that case the deformation parameter $q$ is real, 
not a pure phase.
On the other hand,  
level-rank of Chern-Simons observables 
could have important implications 
for large-$N$ dualities of Chern-Simons theories
since it relates the large-$N$ limit to the large-$K$ limit
of finite $N$ observables.

\section*{Appendix}
\setcounter{equation}{0}
\def\theequation{A.\arabic{equation}}

In this appendix, we summarize various results for finite-dimensional
and affine Lie algebras needed in the main part of the paper.

\vs{.1in}
\noindent{\bf Representations of $\uN$ }
\vs{.1in}

\option
The $\suN$ generators
$T^a$ $(a=1, \ldots, N^2-1)$ 
satisfy
$[T^a, T^b] =  i f^{ab}_{~~c} T^c$,
and the Killing form is defined as $g^{ab} = \Tr (T_\one^a T_\one^b)$,
with the trace taken in the defining representation of $\suN$.
(With this normalization of $g^{ab}$, the roots of $\suN$ have length two.)
A representation $R$ of $\suN$ is characterized 
by a Young tableau with row lengths 
$\ell_i$ $(i=1, \ldots, N-1)$ satisfying
$\ell_1 \ge \ell_2 \ge \cdots \ell_{N-1} \ge 0$,
and with number of boxes 
$r = \sum_{i=1}^{N-1} \ell_i$.
The quadratic Casimir $C_2(R)$ of the 
$\suN$ representation $R$ is given by 
\bea
\sum_{a,b = 1}^{N^2-1}
g_{ab} T_R^a T_R^b 
     & = & C_2(R) \,\1_{\dim R} \\
\label{eq:suncas}
C_2(R) & = & N r + T(R) - {r^2 \over N} \\
T(R) & = & \sum_{i=1}^{N-1} \ell_i (\ell_i - 2i + 1)  
\eea
where $g_{ab}$ is the inverse of $g^{ab}$.

Representations of $\uN = [\suN \times \uone]/\ZN$
are those representations 
$(R,Q)$ of $\suN \times \uone$ 
that satisfy $Q=r$ mod $N$.   
Here, $Q$ is the eigenvalue of the $\uone$ generator $T^0$,
which is normalized so that the defining representation has $Q=1$,
implying that $g^{00} = N$.
The quadratic Casimir of the $\uN$ representation $(R,Q)$ is then
\bea
\sum_{a,b = 0}^{N^2-1}
g_{ab} T_{(R,Q)}^a T_{(R,Q)}^b 
     & = & C_2 (R,Q) \, \1_{\dim R} \\
\label{eq:uncas}
C_2 (R,Q)  & = & C_2(R) +  {Q^2 \over N}
            \ =\   N r + T(R) + 2 s r + N s^2 \\[.2in]
	Q & = & r + N s \,.
\eea
One can characterize the $\uN$ representation $(R,Q)$
by an extended Young tableau $\cR$, 
obtained by prepending $s$ columns of $N$ boxes 
to the Young tableau for $R$,
giving a tableau with row lengths 
$\bell_i = \ell_i + s$ $(i=1, \ldots, N-1)$ 
and $\bell_N = s$,  
and with number of boxes
$Q = \sum_{i=1}^{N} \bell_i$.
Since $s \in \Z$, the $\bell_i$ may be negative, 
in which case one can represent the tableau using ``anti-boxes'';
cf. appendix D of ref.~\cite{Aganagic:2005dh}.
The quadratic Casimir (\ref{eq:uncas}) may be re-expressed as
\bea
\label{eq:CcR}
C_2(\cR) 
&=& 
C_2(R,Q) \, = \,  N Q  + T(\cR) 
\\
\label{eq:TcR}
T(\cR)    
&=& 
\sum_{i=1}^{N} \bell_i (\bell_i - 2i + 1)  \,.
\eea

\vs{.1in}
\noindent{\bf Representations of $\suNK$ and $\uoneKp$ }
\vs{.1in}

\option
Next, we review the representations of the affine Lie algebras
$\suNK$ and $\uoneKp$.

The generators of $\suNK$ satisfy
\be
[J_m^a, J_n^b] =  i f^{ab}_{~~c} J_{m+n}^c +  K g^{ab} m \delta_{m+n}
\ee
where the level, $K$, is integer-valued when $g^{ab}$ is normalized as above.
Integrable highest-weight representations of $\suNK$ 
are characterized by $\suN$ representations $R$ 
whose Young tableaux have $\ell_1 \le K$.
This representation has conformal weight 
\be
\label{eq:sunh}
h_R = { \half C_2(R) \over K+N}
\ee
and affine character 
\be
\label{eq:sunchar}
\chi_R (\tau)= 
\Tr \left( q^{L_0 - c/24}\right) =  q^{h_R - c/24} ( \dim R + \cdots),
\qquad\qquad
q = \e^{2\pi i \tau}
\ee
where the central charge is $c = K(N^2-1)/(K+N)$.
The characters of integrable highest-weight representations
transform into one another under the
modular transformation $\tau \to -1/\tau$:
\be
\label{eq:sunxf}
\chi_R (-1/\tau)= \sum_{R'}  S_{RR'} \, \chi_{R'} (\tau)
\ee
and into themselves, up to a phase, under $\tau \to \tau + 1$:
\be
\chi_R (\tau+1)= \sum_{R'}  T_{RR'} \, \chi_{R'} (\tau),
\qquad  
T_{RR'} = \e^{2 \pi i \left(h_R - {c / 24}\right) } \delta_{RR'}\,.
\ee
Integrable highest-weight representations of $\suNK$ 
may be grouped into cominimal equivalence classes
(or simple-current orbits) of order $N$ (or some divisor of $N$).
These classes are generated by the simple current $\sig$,
which rotates the Dynkin indices of the Dynkin diagram of $\suNK$.
In terms of Young tableaux, $\sig(R)$ is obtained by
adding a row of $K$ boxes to the top of the Young tableau of $R$,
and stripping off any columns of length $N$ that may result.
The conformal weights of cominimally-equivalent representations 
are related by 
\be
\label{eq:hcomin}
h_{\sig(R)} = h_R + { NK -K - 2r \over 2N}
\ee
and the modular transformation matrix\footnote{
The modular matrix $S$ here differs from that given in 
ref.~\cite{Mlawer:1990uv} by complex conjugation.}
obeys 
\be
\label{eq:Scomin}
S_{\sig^n(R) R' } = \e^{2\pi i n r' / N } S_{RR'}
\ee
where $r'$ is the number of boxes of the tableau representing $R'$.

The generators of $\uoneKp$ satisfy
\be
[J_m^0, J_n^0] =  \Kp  m \delta_{m+n} \,.
\ee
Highest-weight representations of this algebra
are labelled by their $J_0^0$ eigenvalue 
$Q$, and have conformal weights
\be
\label{eq:uoneh}
h'_Q = { \half Q^2 \over \Kp }
\ee
and affine characters
\be
\label{eq:uonechar}
\chi'_Q (\tau)  =   {q^{h'_Q} \over \eta(\tau)}
        =   q^{h'_Q - \twentyfourth} \prod_{n=1}^\infty {1\over (1 - q^n)} \,.
\ee

\vfil\break
\providecommand{\href}[2]{#2}\begingroup\raggedright\endgroup

\end{document}